# Electronic behavior in mats of single-walled carbon nanotubes under pressure

R. Falconi, J. A. Azamar,* R. Escudero.

*Instituto de Investigaciones en Materiales, Universidad Nacional Autónoma de México. A. Postal. 70-360. México, D. F. 04510 MEXICO.*

*Departamento de Física Aplicada. Cinvestav, 97310, Mérida, Yuc. México.

## ABSTRACT

*Single-walled carbon nanotubes (SWNTs) have many interesting properties; they may be metallic or semiconducting depending on their diameter and helicity of the graphene sheet. Hydrostatic or quasi-hydrostatic high pressures can probe many electronic features. Resistance - temperature measurements in SWNTs from normal condition and under 0.4 GPa of quasi-hydrostatic pressures reveal a semiconducting-like behavior. From 0.5 to about 2.0 GPa the resistance changes to a Kondo-like feature due to magnetic impurities used to catalyse the nanotube formation. Above 2.0 GPa, they become metallic and at about 2.4 GPa the resistance decreases dramatically around 3 K suggesting a superconducting transition.*

Two unique characteristics of carbon nanotubes are their different electronic behavior depending on diameter and helicity [1, 2]. Electronic and transport properties of single-walled carbon nanotubes (SWNTs) also differ depending on chirality [3-5]. Many theoretical studies indicate that the local density of electronic states changes if the graphene sheet is wrapped in a zigzag or an armchair configuration. Armchair nanotubes are metallic, whilst zigzag one are semiconducting [4]. Recently experiments in very narrow isolated SWNTs show superconductivity [6]. Also, signs of superconductivity in ropes of armchair SWNTs at temperatures below 0.55 K, has been observed [7]. Theoretical studies suggest intertube coupling effects between nanotubes [8] to explain the superconducting behavior observed by Kociak *et al.*, [7] rather than an electronic mechanism as proposed by Tang *et al.*, [6] for narrow nanotubes. In this work we report experimental studies in mats of SWNTs using high pressure produced in a diamond anvil cell (DAC). Interestingly, the samples show different electronic behaviors with the applied pressure: Luttinger liquid, Kondo-like features, and possible superconductivity at low temperature. The results are discussed taking account theoretical predictions.

The SWNT were obtained from MER Corporation, they have average diameter of 0.7 – 3 nm, and lengths from 2 – 20 μm. In order to prepare the samples for transport high-pressure studies they were purified [9] by heated in air to 300 °C for 24 h and dispersed by ultrasound in hydrochloric acid, and refluxed for 4 h at 100 °C. After that, they were washed with distilled water and ethanol (95 % spectrophotometric grade) and filtrated through a membrane filter of 0.20 μm pore size. This method produces a mat-like sample that was dried at 100 °C for about 2 h. In Fig.1 we show TEM pictures of one sample after the cleaning treatment. According to Figure 1**a** the mat-like sample is formed by an interconnected network of ropes of SWNTs and some macroscopic impurities, mainly amorphous carbon particles. Figure 1**b** shows atomic resolution for some of the ropes in the sample. In this figure we also observe black dots that were analyzed to be carbon coated metallic impurities. Those nanotubes from MER contain different amounts of Co, Ni or Fe impurities used to catalyse the formation of the tubes. Once the samples were cleaned with hydrochloric acid was found atomic percentages of about 0.40, 0.12, 0.03 for Co, Ni, and Fe respectively. The TEM analyses revealed that these impurities have different sizes ranging from 1 to 50 nm and are distributed in a inhomogeneous way along the all specimen. Other TEM pictures show that some small metallic particles are situated close or on the surface of the ropes.

To perform the Resistance – Temperature (R-T) measurements, at different pressures, the samples were divided and shaped to appropriate sizes and mounted in the diamond anvil cell. Four gold wires of 10 μm diameter were used for the R-T measurements; they were performed using an Oxford Optistat Cryostat adapted to introduce the DAC. The change of pressure can be adjusted on the top of the cryostat with an oil manometer. The experiments were performed from room temperature to about 1.6 K. The pressure calibration was determined with the change of the superconducting temperature of Pb.

Measurements at atmospheric pressure were determined in a Quantum Design system from 300- 2 K. WDX (Wavelength Dispersive X-ray Detectors)/EDX (Energy Dispersive Spectrometer) analysis was carried out in a Jeol transmission electron microscopy. For the quasi-hydrostatic pressure studies we used Cu-Be gaskets filled with fine powder of MgO < 2 μm size. Additional details on the experimental method are given in Falconi *et al*. [10].

The main panel of Fig. 2 shows the overall behavior of typical R-T curves for one of the samples normalized to R(50 K), at different quasi-hydrostatic pressures. As observed before, such behavior is completely different for that of graphite [11]. Measurements of the R-T characteristic at atmospheric pressure reveal a semiconducting-like behavior which was possible to fit to a $R \sim T^{-\beta}$ law with $\beta \sim 0.66$ (see inset **a** of Fig. 2), where $\beta = (1-g)/2$. These values may indicate a Luttinger behavior with parameter $g \sim 0.2$, well below of the Fermi liquid value. This Luttinger parameter was calculated (see Ref. [1]) using appropriate values for graphite; Fermi velocity $v_F = 8 \times 10^5$ m/s and screening length about 1000 Å, giving $g \approx 0.2$. Different sets of mat-like samples give values for $\beta$ around to 0.6 at room pressure conditions. We point out that in all our different mats of

SWNTs the behavior follows this law, instead of the found in other study where was observed a two dimension variable range hopping [12]. Shiraishi *et al*., [13] have also observed a Luttinger liquid characteristic in single-walled carbon nanotubes networks. It has been point out that the Luttinger law persists with the applied pressure [14]. However, as we will see below, the overall trend changes as soon as the external pressure is applied. However, we remark that the electronic processes that we found as a function of temperature in mats reflects qualitatively at least the one-dimensional nature of its components [15]. Different processes can change the electronic characteristics of mats with pressure; i) nanotube deformation, ii) stretching and squeezing by random intertubular contact giving rise to a interlinked networks of nanotubes, iii) change of bonding from $sp^2$ to a $sp^3$ character, due to change of curvature. SWNTs will be squeezed in different degree depending how compact and oriented is the mat, the electronic conduction process will changes with pressure because bonding changes with applied pressure

All our electrical resistance measurements show in general three main trends of behavior: i) from atmospheric to low pressures (about 0.2-0.3 GPa) R-T shows a semiconducting-like behavior; ii) additional increase of applied pressure to about 2.0 GPa changes the semiconducting-like behavior to a Kondo-like feature, the R-T curve shows a minimum, and a logarithmic increase at low temperature (the minimum starts at around 17 K and decreases with pressure, this is shown in Fig 2, inset **b**), iii) at pressure above 2.4 GPa, R-T characteristic suffers a dramatic change to metallic behavior, and finally at about 3 K the resistance drops precipitously, as occurring in a superconducting transition (see Fig. 3).

The analysis of the Kondo feature (the minimum in R-T and the logarithmic up-turn) evolves and decreases with pressure, becomes shallower, and disappears at about 2.4 GPa. A fit of the low temperature electrical resistance to a logarithmic law confirms the Kondo nature of this feature. We believe that the influences of magnetic impurities on the conduction electron spins strongly affect the electrical transport of SWNTs, and leads to anomalous behavior in the R-T characteristic [16,17]. The Kondo effect results from the interaction between the magnetic moment of the transition metal and the spin of the conduction π electrons of the nanotubes [17]. T. Odom *et al*., [16] report scanning tunnelling experiments and show that Co cluster on semiconducting SWNT exhibited no feature at $E_F$ related to a Kondo resonance peak, and that this peak emerges when the SWNT is metallic. The authors claim that these facts emphasize the necessity of conduction electrons to interact with the magnetic cluster in order to observe the Kondo feature. In our case the samples (with magnetic impurities) behave as a semiconductor like at room pressure but become metallic with the applied pressure, and the Kondo feature emerges.

Inset **b** of Fig 2 presents the decreasing of the Kondo temperature minimum with pressure. This is explained considering the Fiete *et al*., model [18], they found that ferromagnetic Co nanoclusters on metallic nanotubes tends to suppress the Kondo temperature; thus, for Co cluster the itinerant model leads to antiferromagnetic coupling between nanotubes and cluster spins. Experimentally the effect of pressure is to bring together small Co antiferromagnetic clusters, the resulting effect is then that nanoclusters change to a ferromagnetic state decreasing therefore the Kondo temperature [18]. However, we do not understand however which could be the role of the macroscopic metallic particles in the pressure induced elecronic behavior.

If pressure is further increased the temperature of the resistance minimum tends to disappear, and the R(T) characteristic becomes metallic. As soon as this change occurs a dramatic reduction of R(T) is produced. The precipitously reduction of resistance without doubt can be seen as a superconducting transition. However in our experiments never zero resistance was reached. It is worth mentioning that the minimum values of resistance reached for some of our samples mats (in terms of normalized resistance values) were 0.03 at about 3 GPa.

An interesting fact observed in all our measurements is that the sample resistance at high temperatures always increases as the pressure is increased. Fig. 4 shows a plot of the general trend of R(P), taken for example at 115 K for two samples. There, we have plotted, for comparison, the percentage of change of $\Delta R/Ro$ versus pressure. The trend indicates that resistance at high pressures always increases, with similar rates of change in both samples. Experimentally we found that the effect is reversible until a certain pressure, beyond that point, the resistance follows increasing as pressure increases. If the pressure is released the resistance drops slightly but without coming back to the initial value. Theoretical studies [19] show that in the elastic regime, and depending on diameter, a radial deformation in zig-zag nanotubes closes the band gap at the Fermi surface. In others zig zag SWNTs, the band gap initially increases and then decreases when strain exceeds certain threshold value, eventually all zig zag tubes will suffer an insulator-metal transition in that regime. In contrast, in metallic nanotubes a radial deformation decreases the density of states at the Fermi level and a metal - semiconductor transition takes place [20,21]. Other works show that beyond a critical deformation several carbon atoms become $sp^3$ coordinated, in such a way that $\pi$ delocalized electrons become $\sigma$ localized states [22]. The nanotubes conforming the ropes can be elliptically deformed by pressure-induced interlinking [23], thus tending to form covalent bonds with a $sp^3$ character between them at points of higher curvature by hybridization. This effect is bigger for closed nanotubes [24]. In our case, the fact that the resistance as a function of pressure increases at room pressure and the pressure induced metallic transition could be associated with a competence between nanotubes with different chirality. However, we do not discard defects due to cleaning chemical procedures may affect the transport properties.

Lastly, we noted that in different sets of measurements the metallic-like transition occurs at temperatures fairly high. Interesting enough, the transition to the metallic regime and to the superconducting-like behavior is quite wide, which is a typical effect for a low dimensional system [25], at the moment we have not a clear explanation for this different behavior, and for the superconductivity. However, one effect to be considered is the different amount of magnetic impurities in contact with nanotubes. [16].

In summary the electronic behavior of mats of carbon nanotubes under pressure are reported. They show a complex electronic behaviour. At low pressure resistance R(T) behaves as 1-d system, with a Luttinger liquid characteristic. With increasing pressure R(T) presents a minimum and a logarithmic growth at low temperature, typical of Kondo behavior, due to magnetic impurities present in the nanotubes used in the catalyst process. At higher pressures the Kondo feature disappears and a metallic characteristic develops. Further increase of pressure changes the metallic behavior to a superconducting-like transition.

**Acknowledgements**

Supported by UNAM-DGAPA grant IN102101, and by CONACyT-MEXICO grant G0017, and 38490E. We thank F. Morales, R. Rangel, M. de Llano and F. Silvar for technical assistance and discussions.

**Fig. 1**. High-resolution TEM images of purified SWNTs. **a**). Shows a picture of the mat-like sample. **b**). Shows a picture with atomic resolution of some nanotube ropes. The black dots are carbon coated particles of Co, Ni, or Fe.

**Fig. 2.** Normalized resistance vs temperature (R-T) measurements at different quasi-hydrostatic pressures in mats of closed SWNTs. (**a**) R-T curve from 300 to 1.6 K at atmospheric pressure. At low pressures overall behavior is semiconducting-type; as soon as pressure increases from 0.2 to 2.0 GPa, the trend changes to a type of Kondo-like behavior and the resistance minimum decreases from about 17 K to about 2 K. (**b**) shows the decreasing temperature of the resistance minimum as function of pressure. This decreases as pressure increases until it disappears. In this particular set of measurements the minimum is reduced at a rate of -6.7 K/GPa. We found this same general trend in other measurements.

**Fig. 3.** Superconducting like transition in the R(T)/R(50K) vs T curve at pressure of 2.4 GPa from 120 K to low temperature in SWNTs purified with hydrochloric acid, but still with a small amount of magnetic impurities of transition metals.

**Fig. 4.** Percentage of change in resistance ($\Delta R/R_0$) with pressure at 115 K for samples with different purification process. $R_0$ is the resistance at the lower pressure.

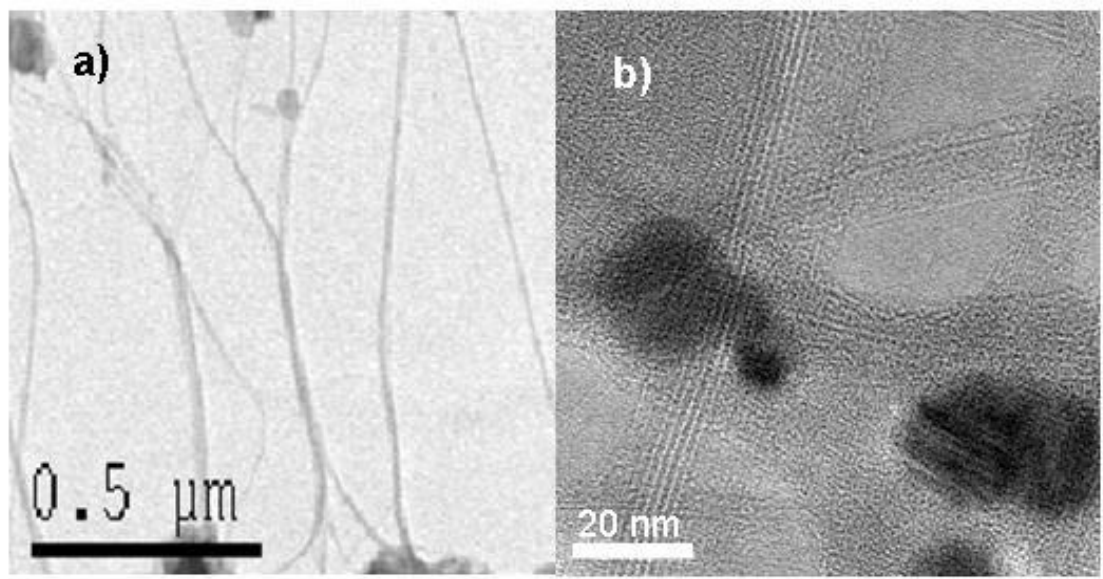

**Figura 1**

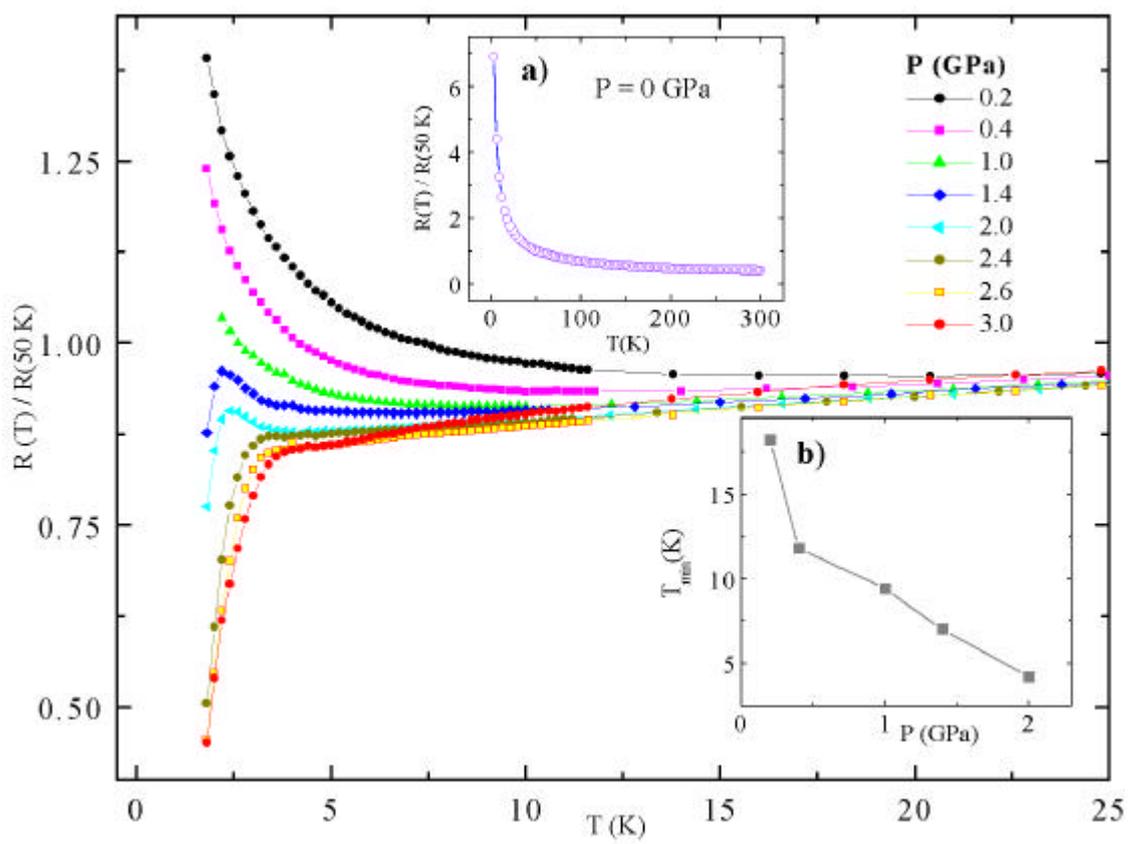

**Figura 2**

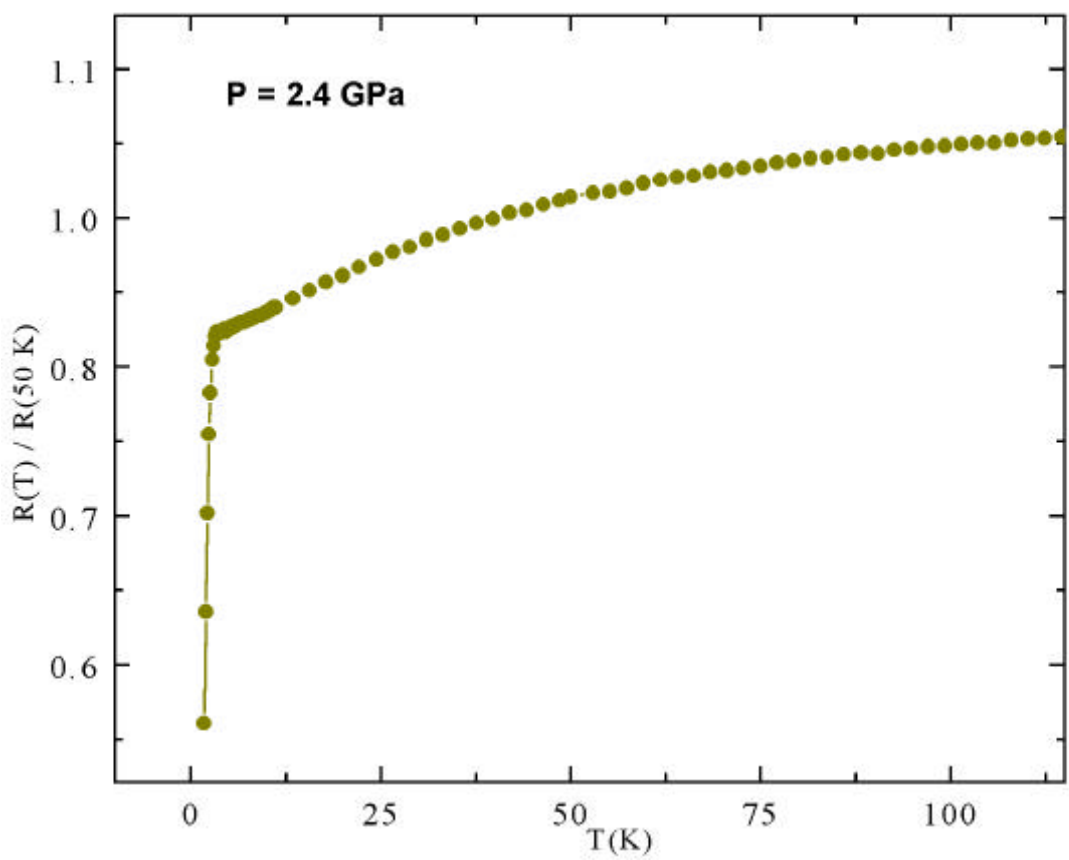

**Figura 3**

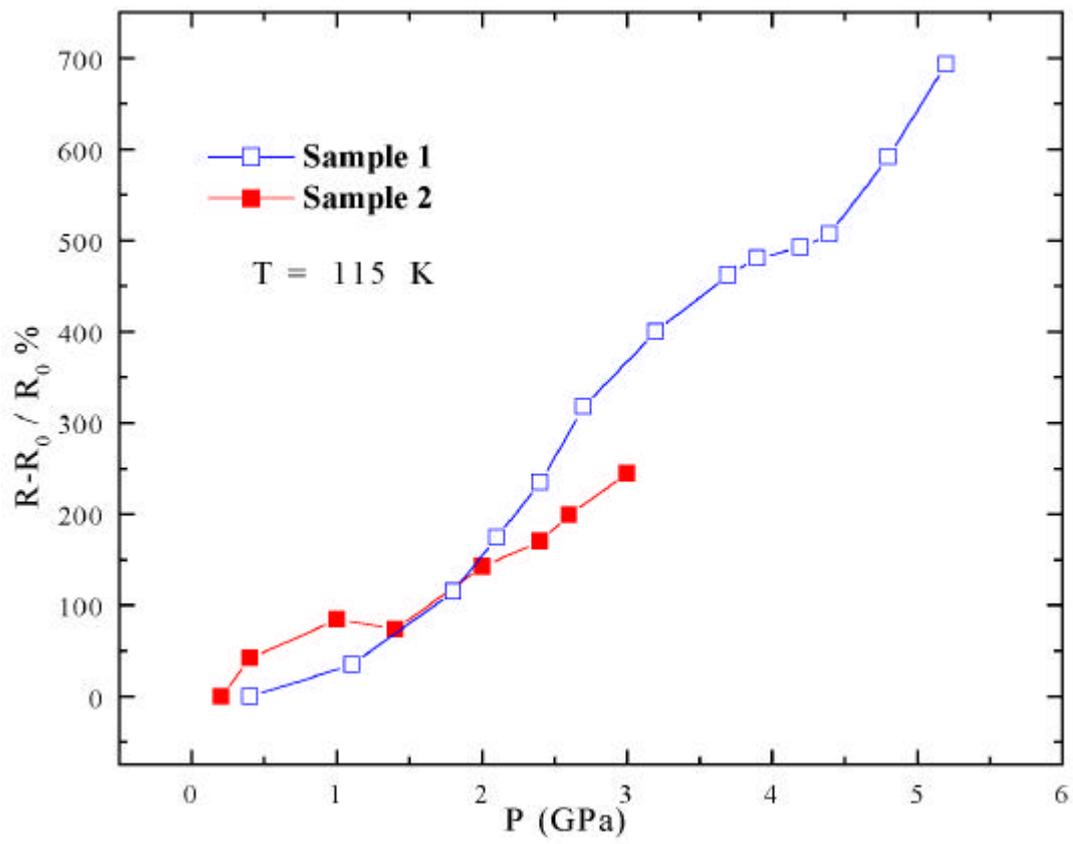

**Figura 4**